\documentclass[useAMS,usenatbib]{mnras}
\usepackage{times,graphics,graphicx,multicol,amssymb} 

\def\aap{A\&A}

\def\apj{ApJ}

\def\apjl{ApJ}
\def\mnras{MNRAS}
\def\araa{ARA\&A}

\def\nat{Nat}
\def\aaps{A\&A Supp.}
 
\def\apjs{ApJS}

\def\ssr{Space Sci. Rev. }

\def\lesssim{\mathrel{\hbox{\rlap{\hbox{\lower4pt\hbox{$\sim$}}}\hbox{$<$}}}}
\def\gesssim{\mathrel{\hbox{\rlap{\hbox{\lower4pt\hbox{$\sim$}}}\hbox{$>$}}}}

\def\chandra    {{\em Chandra}\/}

\begin{document}

\author[Morandi et al.]
{Andrea Morandi${}^1$\thanks{E-mail: andrea.morandi@uah.edu}, Ming Sun${}^1$\thanks{E-mail: ms0071@uah.edu}, John Mulchaey${}^2$, Daisuke Nagai${}^{3,4,5}$, Massimiliano Bonamente${}^1$\\
$^{1}$ Physics Department, University of Alabama in Huntsville, Huntsville, AL 35899, USA\\
$^{2}$ The Observatories of the Carnegie Institution for Science, Pasadena, CA 91101, USA\\
$^{3}$ Department of Physics, Yale University, New Haven, CT 06520, USA\\
$^{4}$ Department of Astronomy, Yale University, New Haven, CT 06520, USA\\
$^{5}$ Yale Center for Astronomy \& Astrophysics, Yale University, New Haven, CT 06520, USA\\
}

\title[The galaxy group NGC~2563]
{Gas distribution and clumpiness in the galaxy group NGC~2563}
\maketitle

\begin{abstract}
We present a \chandra\ study of the hot intragroup medium (hIGM) of the galaxy group NCG2563. The \chandra\ mosaic observations, with a total exposure time of $\sim$ 430 ks, allow the gas density to be detected beyond $R_{200}$ and the gas temperature out to 0.75 $R_{200}$. This represents the first observational measurement of the physical properties of a poor groups beyond $R_{500}$. By capitalizing on the exquisite spatial resolution of \chandra\ that is capable to remove unrelated emission from point sources and substructures, we are able to radially constrain the inhomogeneities of gas (``clumpiness''), gas fraction, temperature and entropy distribution. Although there is some uncertainty in the measurements, we find evidences of gas clumping in the virialization region, with clumping factor of about 2 - 3 at $R_{200}$. The gas clumping-corrected gas fraction is significantly lower than the cosmological baryon budget. These results may indicate a larger impact of the gas inhomogeneities with respect to the prediction from hydrodynamic numerical simulations, and we discuss possible explanations for our findings.
\end{abstract}


\section{Introduction}\label{intro}
Galaxy clusters represent the last stage of the hierarchical large-scale structure formation, and they are powerful laboratories for astrophysics, cosmological parameters and testing models of structure formation. In particular, galaxy groups are less massive, gravitationally bound systems than galaxy clusters, with typical hIGM temperature less than 2 keV \citep{ponman1996,mulchaey1996,mulchaey2000,eke2004}. Because of shallow gravitational potential, galaxy groups are indeed not simply scaled-down version of more-massive clusters \citep{ponman1999,ponman2003,sun2009}. While the physics of the hIGM of latter system is gravity-dominated, groups are systems where the roles of complex baryon physics (e.g. cooling, galactic winds and AGN feedback) are much more pronounced. Thus, they are particularly useful for study of the physics of these non-gravitational processes, and to constrain cluster formation models \citep{ponman1999,bower2001,ponman2003,voit2005a,sun2012}.

In particular, X-ray observations provide the best means for studying emission from the hot gas in groups. Studying the hIGM provides stringent constraint on the underlying DM distribution, since the hIGM closely traces the group potential well; moreover, the hydrostatic equilibrium (HE) assumption should be robust at least within $R_{500}$\footnote{$R_{500}$ corresponds to the radius enclosing an overdensity of 500 with respect to the critical density of the Universe.}, allowing the gravitating mass to be derived directly from the X-ray observables \citep{sarazin1988}. Of particular interest in X-ray are the gas density, temperature, entropy $K =  kT/n_e^{2/3}$, which allow to infer the cluster masses (under the assumption of HE), as well as to infer the amount of non-gravitational feedback that affects the properties of the ICM.

\begin{figure*}
\begin{center}
\includegraphics[scale=0.75]{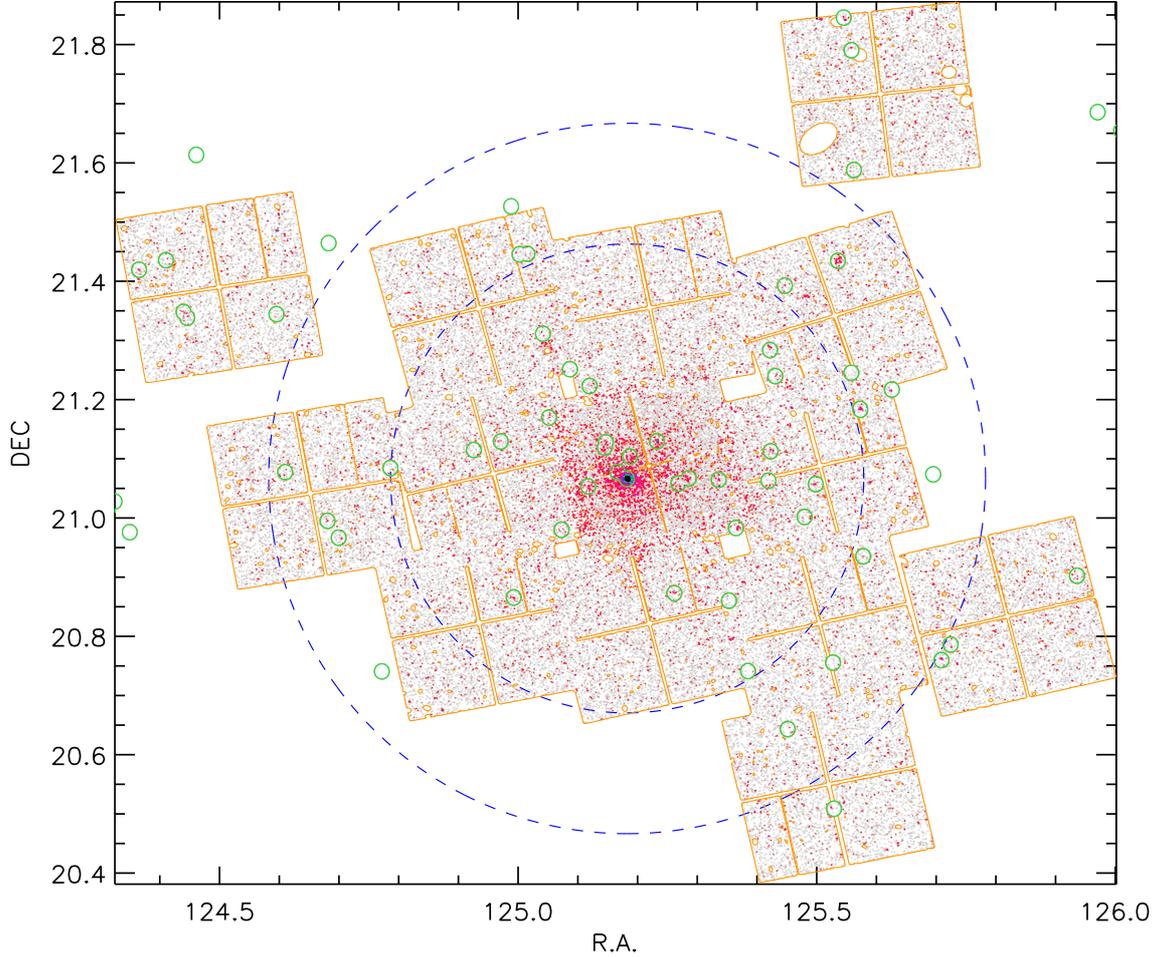}
\caption[]{\chandra\ mosaic image of NGC~2563 (with a pixel size of 0.32~kpc). The blue circles indicate $R_{500}$ and $R_{200}$, respectively ($R_{500}=456\pm28$~kpc and $R_{200}=691\pm44$~kpc, as we measured in \S\ref{datasets3}), while the orange circles indicate the masked point sources and substructures. We also show the 82 optical group members (\S\ref{opt1}) in green circles, where the radius corresponds to a size of 15~kpc. Note that a few optical group members are not visualized since they fall outside the boundaries of the image. The fractional coverage of the virialization region ($R_{500}-R_{200}$) is $\sim0.5$ for our mosaic observations, while for $R\lesssim R_{500}$ is 0.94.}
\label{eft35n3fr}
\end{center}
\end{figure*}

Recently, a lot of attention has been polarized on the study of the virialization region of clusters, which is the front line of cluster formation, and an important to test the validity of the CDM large-scale formation model. Since clusters are still forming in the present epoch, non-equilibrium phenomena are expected in the outskirts, including deviations from the hydrostatic approximation, multi-temperature distribution, non-equipartition and gas inhomogeneities (``clumpiness''). In particular, the gas clumping factor $C=<n_e^2>/<n_e>^2$ has been observationally bracketed in the range 1-2 \citep{morandi2013b,morandi2014,eckert2015}, in agreement with predictions from hydrodynamic simulations \citep{nagai2011,battaglia2013}. Other studies, however, found non-negligible larger values of the gas clumping, in tension with the prediction of the CDM model \citep{walker2013}. Most of these studies focused on massive objects, while only a very few on groups. It is thus important to extend these studies to lower-mass clusters, where the physics of non-gravitation processes has larger ramifications - with respect to massive clusters - on the properties of the hIGM for both astrophysics and cosmology. To date, most observations of the gas in groups have been restricted to within $\sim R_{2500}-R_{500}$, while regions beyond $R_{500}$ in groups remain little explored \citep[e.g.][]{sun2012} and would require deep observations for a single group. Only a very few analyses have been able to study groups out to the virial radius \citep[e.g.][]{su2015,buote2016,wong2016}. However, no system with global temperature $kT\lesssim1.4$ keV has been studied beyond $R_{500}$ \citep[][]{sun2009,sun2012}.

In the present study, we present an analysis of the hIGM physical properties, namely gas density and gas clumping, gas temperature distribution, entropy and gas fraction, of the galaxy group NGC~2563. This group appears a relaxed group in X-rays, with no significant sign of disturbances. Its X-ray emission peak coincides with the position of the Brightest Central Galaxy NGC~2563, as expected for a reasonably relaxed group. This is also a cool core (CC) group \citep{johnson2009}. Like other CC systems, NGC~2563 shows a strong spike in the X-ray surface brightness profile and a drop in the temperature of about a factor of two in the central region. The cool-core corrected X-ray temperature (in the radial range $(0.15-0.75)R_{500}$ ) is $T_{\rm X}=1.20\pm0.04$~keV from our analysis. The X-ray physical properties of NGC~2563 are summarized in Table \ref{tabdon}. This system is characterized by 15 mosaic \chandra\ observations to reach beyond $R_{200}$, allowing the virialization region to be studied.

The paper is organized as follows. In \S\ref{satecn} we present the cluster dataset and we describe the X-ray analysis. Optical properties are discussed in \S\ref{opt1}. The results on the X-ray physical parameters, including gas temperature, density and gas inhomogeneities are presented in \S\ref{phys5346}. The results on the gas and baryon fractions are presented in \S5. In \S\ref{conclusion33} we summarize our conclusions. Throughout this work we assume the flat $\Lambda$CDM model, with matter density parameter $\Omega_{\rm m}=0.3$, cosmological constant density parameter $\Omega_\Lambda=0.7$, and Hubble constant $H_{0}=100h \,{\rm km\; s^{-1}\; Mpc^{-1}}$ where $h=0.7$. One arcminute corresponds to 19.2 kpc at the NGC~2563 redshift of 0.0157. Unless otherwise stated, we report the errors at the 68.3\% confidence level.

\section{Data reduction and analysis}\label{satecn}

\begin{table}
\begin{center}
\caption{Properties of galaxy group NGC~2563. We report the redshift $z$, the optical velocity dispersion \protect\citep{rasmussen2012}, the cool-core corrected X-ray temperature (in the range $(0.15-0.75)R_{500}$ ) $T_{\rm X}$, the emission-weighted abundance is $Z$ (in the range $(0.15-0.75)R_{500}$ and in solar units $Z_{\odot}$), and the bolometric X-ray luminosity $L_X$ within $R_{500}$.}
\begin{tabular}{l@{\hspace{1.5em}} c@{\hspace{1.5em}} c@{\hspace{1.5em}} c@{\hspace{1.5em}} c@{\hspace{1.5em}} }
\hline \\
$ z $ & $\sigma_v$ & $T_{\rm X}$ & $Z$ & $L_X$ \\
      & km/s &    (keV)     &  ($Z_{\odot}$) & ($10^{43}$erg/s)\\
\hline \\
0.0157 & $364^{+36}_{-33}$  & $1.2\pm0.04$ &  $0.16\pm0.04$ & $1.0\pm0.1$\\
\hline \\ 
\end{tabular}
\label{tabdon}
\end{center}
\end{table} 

\subsection{\chandra\ observations}\label{datasets}
NGC~2563 has been covered by 15 \chandra\ ACIS-I observations in 15 mosaic positions in 2007 - 2009 (PI: Mulchaey, see Table \ref{tabdon2}), with a total exposure time of $\sim430$ ks. This mosaic configuration was originally chosen in order to track the distribution of group members in NGC~2563, via a $3\times 3$ central mosaic to cover the inner $45' \times 45'$ of the group and five outer pointings to cover the virialization region and beyond.

The data reduction was carried out using the CIAO 4.8.1 and Heasoft 6.19 software suites, in conjunction with the \chandra\ calibration database (CALDB) version 4.7.2. Here we briefly summarize aspects of the data preparation and analysis of NGC~2563, while we refer to \cite{morandi2013b} for a further description of the X-ray analysis. Briefly, the data were reprocessed from the 'level 1' event files, in order to produce a 'level 2' file. Periods of high background were identified via the task {\tt deflare} in the light curve (created via the tool {\tt dmextract}) from a low surface brightness region of the CCDs, not contaminated by the source emission. Data from these intervals were thus excised. Point sources were detected in the 0.7-7.0 keV image with the {\tt wavdetect} CIAO task, which was supplied a weighted exposure map by assuming a powerlaw spectra with slope of 1.4. The detection threshold ($10^{-6}$) guaranteed $\sim 1$ spurious source per CCD. We also run {\tt wavdetect} with wavelet scales in the range 1-64 pixels on the images, with each succeeding scale value being a factor of $\sqrt{2}$ larger than the previous one. All detected sources were confirmed visually.

\subsection{Spatial analysis}\label{datasets2}
For the spatial (or surface brightness) analysis, our goal is to measure the inhomogeneities of the surface brightness distribution, and hence gas density. We produced X-ray images from the level-2 event file in the energy range 0.7-2.0 keV. This energy band was chosen in order to: i) maximize the S/N in the outskirts of the clusters, and ii) minimize the dependency of the cooling function on the temperature and metallicity, which can be otherwise non-negligible at the lower temperatures of galaxy groups. First, we determined the centroid of the brightness image by inferring the position where the brightness derivatives along two orthogonal directions become zero (R.A. (J2000)=125.149 degrees, Decl. (J2000)=21.0678 degrees). This method is particularly suitable for NGC~2563 over other methods (e.g. brightness-weighted centroid). Our method is indeed insensitive to the presence of neighboring structures. We checked that the centroid of the surface brightness is consistent with the center of the BCG, the shift between them being evaluated to be a few arcsec: note that the uncertainty on the X-ray centroid estimate is comparable to the applied rebinning scale on the X-ray images.

\begin{table}
\begin{center}
\caption{Particulars of the NGC~2563 observations, including the observation ID, the observation mode and the effective exposure time (ks). The total exposure time is $\sim430$~ks.}
\begin{tabular}{l@{\hspace{1.8em}} c@{\hspace{1.8em}}  c@{\hspace{1.8em}}  }
\hline
ID & MODE & Effective Exp.   \\
   &           &  {(ks)\quad } \\
\hline \\
7925 & VFAINT &   48 \\
7926 & VFAINT &   29 \\
7927 & VFAINT &   30 \\
7928 & VFAINT &   30 \\
7929 & VFAINT &   28 \\
7930 & VFAINT &   28 \\
7931 & VFAINT &   30 \\
7932 & VFAINT &   30 \\
7933 & VFAINT &   30 \\
7934 & VFAINT &   19 \\
7935 & VFAINT &   31 \\
7936 & VFAINT &   27 \\
7937 & VFAINT &   29 \\
7941 & VFAINT &   30 \\
9804 & VFAINT &   12 \\
\hline        
\label{enf}   
\end{tabular} 
\label{tabdon2}
\end{center}
\end{table}   

Next, we created an exposure-corrected image from a set of observations using the \texttt{merge\_obs} to combine the ACIS--I observations (see Figure \ref{eft35n3fr}).
 
We then corrected the images by the exposure maps to remove the vignetting effects. In a massive cluster, where the gas temperature is sufficiently high, customarily one uses the average cluster temperature in order to infer a weighted instrument map which is weighted according to a specific model, namely the APEC emissivity model \citep{foster2012}, for the incident spectrum. This approach is too simplified in the case of a group (with radial temperature profiles of the order of 1 keV or less), given the stronger dependency of the gas cooling function on the gas temperature and metallicity. Rather, we divide the images in a set of concentric annuli, and calculate a weighted exposure map for each annulus, with temperature and metallicity which enter in the emissivity model as measured in the reference annulus. This provides a vignetting correction which properly accounts for metallicity and temperature radial gradient.

Finally, we measured the flattening and orientation of the X-ray surface brightness. In this respect, the final goal is to perform a comparison between the cosmic filament in X-ray (as probed by the elongation of the X-ray isophotes) and in optical (as probed by the galaxy distribution, see \S\ref{opt1}). We computed the moments of the surface brightness within a circular region of radius $R_{500}$ centered on the centroid of the X-ray image (see \citealt{morandi2010a} for further details on this method). This allows us to estimate the hIGM eccentricity on the plane of the sky and the orientation (position angle) of an elliptical X-ray surface brightness distribution.  We found that the X-ray surface brightness distribution appears to be slightly elongated, with an axial ratio $1.05\pm0.01$ and position angle $93\pm1$ degrees (measured north through east in celestial coordinates). Errors have been calculated via Poisson randomization of the X-ray surface brightness. We use this position angle as a proxy of the direction of the large-scale filament (see discussion in \S\ref{opt1}). 

\subsection{Spectral analysis}\label{datasets3}
We extracted spectra in radial annuli from circular annuli around the centroid of the surface brightness and by using the {\em CIAO} \texttt{specextract} tool from each observation. Individual observations are jointly fitted taking into account the response of each observation, through an absorbed thermal emission model in the energy range 0.7-5 keV. We used a custom algorithm to perform the spectral fit and recover the three-dimensional (3D) or deprojected temperature profile.

The package implements a spectral model $Mod$, with a single-temperature APEC model \citep{foster2012}. The basic physical assumption is thus that the gas temperature is uniform within spherical shells whose radii correspond to the annuli used to extract the spectra. Given this assumption one constructs a model for each annular spectrum that is a linear volume-weighted combination of shell models. We assume that the cluster is spherically symmetric, and it has an onion--like structure with $n$ concentric spherical shells, each characterized by uniform and temperature distribution within it. Therefore, the cluster image in projection is divided into $n$ rings (or annuli) of area ${\bf A}=(A_1,A_2,...,A_n)$. We used the convention that the annulus numbering starts from the inner radius at 0, and they are assumed to have the same radii of the 3D spherical shells of radius ${\bf r}=R_i,\; i=1,...,n$. Hereafter we assume that the index $j$ ($i$) indicates the shell (ring) defined by two radii $(r_{\rm in},r_{\rm out})$. The X-ray boundary (the most external ring) is the $n$th ring with radius $R_n$. It is not required that the annuli include the center but they must be contiguous between the inner and outer radii.

Given a spectral model $Mod[j]$ (i.e. a single-temperature APEC) for each shell $j$ and under the assumption of spherical symmetry and onion--like structure, we assume that the observed image counts are proportional to the sum of volume emission density over $j >= i$ of $vol[j,i] * Mod[j]$. Once the composite source models for each dataset are created the fit analysis can begin. First, we fit the outside shell model using the outer annulus spectrum. We then propagate the model solution for the outside shell to the inward annuli (the so-called 'edge effect'), and fit the next inward annulus / shell jointly for all datasets and until all the shell parameters have been determined. Volumes beyond the boundary radius has been accounted for by assuming a smooth cubic spline extrapolation of gas temperature and density (in logarithmic scale for the latter). We used cross-validation to estimate the smoothing parameter \citep{bouchet1995}.

Free parameters in the spectral model $Mod$ are temperature, metal abundance and normalization. We fixed the hydrogen column density $N_H$ to the Galactic value by using the Leiden/Argentine/Bonn (LAB) HI-survey \citep{kalberla2005}. The redshift was frozen to the value obtained from optical spectroscopy. We adopted the AtomDB (version 3.0.3) database of atomic data\footnote{In the appendix we present a comparison between spectral temperatures and metallicities recovered via the AtomDB databases 3.0.3, 2.0.2 and 1.3.1.}, and we employed the solar abundance ratios from \cite{asplund2009}. We also used the Tuebingen-Boulder absorption model (\texttt{tbabs}) for X-ray absorption by the ISM. We then group photons into bins of at least 20 counts per energy channel and applying the $\chi^2$-statistics. Results of the deprojected temperature and metallicity profiles are presented in Figure \ref{entps3xkn3fr}, along with the projected measurements. For the latter, we used the XSPEC package \citep[][version 12.9.0u]{1996ASPC..101...17A} to perform the spectral fit.

Note that we calculated the mean molecular weight $\mu$ and the protons-to-electrons ratio $n_p/n_e$ for the assumed solar abundance ratios as a function of the measured abundances in the radial annuli. While the dependency upon the metallicity is modest, for reference we quote $\mu$ and $n_p/n_e$ assuming an abundance of 0.3 solar: $\mu=0.60$ and $n_p/n_e=0.85$ by employing the solar abundance ratios from \cite{asplund2009}.

The background spectra have been extracted from regions of the same exposure for the ACIS--I observations, for which we always have some areas ($\gesssim R_{100}$) free from source emission. We refer to \cite{morandi2015} for further details on the modeling of the background. We applied a direct subtraction of the cosmic X-ray background (CXB)+particle background from the measured signal. In this respect, we modeled the soft CXB component by an absorbed power law with index 1.4 and two thermal components at zero redshift, one unabsorbed component with a fixed temperature of 0.1 keV and another absorbed component with a temperature derived from spectral fits \citep[$kT=0.26\pm0.01$ keV, see also][]{sun2009}. We verified a good agreement between the 0.47-1.21 keV observed flux surface density of the soft CXB from the \chandra\ data v.s. the {\em ROSAT} R45 flux \citep[see][for further discussion on the expected correlation]{sun2009}.

The boundary radius of the X-ray surface brightness analysis corresponds to a ratio of source to background flux $\sim$ 15-40 percent at $R_{200}$. Thus, a careful analysis of the systematics and how these uncertainties propagate into the determination of the physical parameters is required.

We point out that we also determined the global temperature $T_X$ in the radial range $0.15-0.75\; R_{500}$. As for $R_{500}$, we infer this quantity by exploiting the existence of low-scatter scaling relations between the total thermal energy $Y_X=M_{\rm gas}\, T_{X}$ and the total mass, as predicted by self-similar theory and high-resolution cosmological simulations \citep{kravtsov2006}. In particular, we used the $Y_{X}-M$ relation measured for the \citet{sun2009} sample of clusters and groups. Note that we do not infer $R_{500}$ by applying HE by means of gas density and deprojected temperature, since we are unable to measure the latter quantity out to $R_{200}$. Moreover, we assume an NFW distribution with parameters $(c,R_{200})$, with the concentration parameter via the $c-M_{\rm vir}$ relation of \cite{dutton2014} based on the results from numerical simulations. This provides an estimate of $R_{200}$ for given value of $R_{500}$, with $R_{500}\simeq 0.66 R_{200}$. A comparison with the HE mass is presented in \S\ref{appe461r23rfc}.

Adding these priors from N-body simulations, e.g. on the concentration parameter and DM density profile, allows us to measure e.g. the gas fraction from the gas density profiles out to $R_{200}$:
\begin{equation}
f_{gas,\Delta}=\frac{M_{gas,\Delta}}{M_\Delta} =\frac{3}{\Delta\rho_{c,z}}\int_0^1\rho_{gas}(x)x^2\, dx.
\label{fgas235}
\end{equation}
with $x={R}/{R_{\Delta}}$, and $\Delta$ being the (fixed) overdensity contrast with respect to the critical density of the Universe.

Lastly, we compare our spectral results with those from previous studies \citep{helsdon2000,finoguenov2007,gastaldello2007}. After the difference on AtomDB and abundance table is accounted for, our results are consistent with these previous results from {\em XMM} and {\em ROSAT}.

\subsection{Clumping analysis}\label{datasets4}
\begin{figure}
\begin{center}
\includegraphics[scale=0.45]{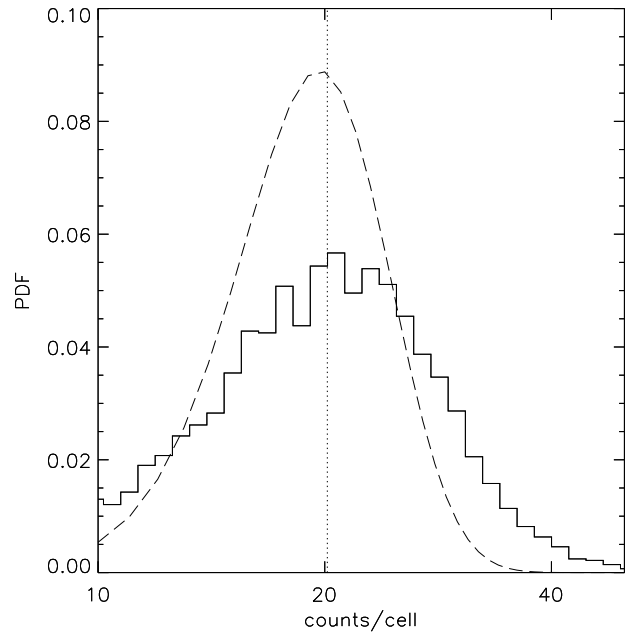}
\caption{Probability density distribution (pdf) of the X-ray surface brightness (counts per cell) in the virialization region $R_{500}-R_{200}$. We rebinned the mosaic image by a factor of 32, that is 1 cell $\sim16$ arcsec, and calculated the corresponding (source+background) X-ray surface brightness pdf (histogram style plot in solid line). The average surface brightness is represented by the vertical dotted line. The long-dashed line represents the theoretical Poisson distribution under the assumption that $S_X$ in the aforementioned region is homogeneous. Note the disagreement between observed and theoretically-expected Poisson distribution, that is the measured $S_X$ pdf is larger due to intrinsic scatter atop of the Poisson fluctuations of the signal. This intrinsic scatter of $S_X$ retains the fingerprint of the 3D gas density inhomogeneities \protect\citep{morandi2013b,morandi2014,eckert2015}.
% We point out that the measured $S_X$ signal is largely dominated by the background in the aforementioned region, that is .
}
\label{entps3xee2bvf}
\end{center}
\end{figure}

As for the gas inhomogeneities, we used a custom package to facilitate deprojection of two-dimensional surface brightness images, in order to recover the 3D gas inhomogeneities. The X-ray package allows inferring {\em directly} the gas density and gas inhomogeneities from the intrinsic fluctuations of the X-ray surface brightness \citep[see][for further discussion]{morandi2013b}. In particular, for the gas clumping, the general idea is to infer gas density inhomogeneities from the observed two-dimensional intrinsic fluctuations of the X-ray brightness. Indeed, an inhomogeneous 3D gas density distribution leaves its imprints on the 2D surface brightness distribution ${\bf \sf S_X}$ atop of Poisson statistical fluctuations and X-ray background (see Fig. \ref{entps3xee2bvf} for visualization purposes). In practice, the algorithm has the ability to disentangle these components and to reverse-engineer the 3D gas density inhomogeneities which generates the measured two-dimensional surface brightness fluctuations.

Note that the fractional coverage of the virialization region ($R_{500}-R_{200}$) is $\sim0.5$ for our mosaic observations, which significantly improves the coverages of observations in preferential directions (e.g. along filaments) carried along narrow arms \citep[e.g.,][]{urban2011,wong2016}.

\section{Group Members and Optical Properties}\label{opt1}
We identified member galaxies of the NGC~2563 group via the NASA/IPAC Extragalactic Database (NED) and the previous work by \citet{rasmussen2012}. From the measured redshifts, we determined group membership by using all galaxies with known recessional velocities within $\pm4\sigma_v$ of the group mean, $\sigma_v\simeq364$ km/s being the the group radial velocity dispersion assuming that the group is virialized \citep{rasmussen2012}. Objects outside the group turnaround radius $R_t$, i.e. the radius where the Hubble flow balances the infall motion, were discarded. This technique identified 82(48) group members within a projected radius of $R_t\approx 2R_{200} = 72'$($R_{200}\sim36'$) from the group center. Out these possible group members, a total of 17 confirmed group members were detected via {\tt wavdetect} \citep[see also][]{rasmussen2012}, including the largest elliptical, NGC~2563 itself, leaving a detectable fingerprint on the main group X-ray emission (see Figure \ref{eft35n3fr}). This galaxy is coincident with the group center as defined by the peak of the diffuse group X-ray emission.

A detailed X-ray analysis of the group members was presented in \cite{rasmussen2012}. For the purpose of the present work, group members have been masked out by conservatively assuming a fixed galaxy radius of 15 kpc, to avoid that X-ray galaxy emission could contaminate the group hIGM emission. We emphasize that this aperture is a very conservative choice as the typical thermal coronae in groups and clusters are much smaller \citep{sun2007,jeltema2008}.

We finally calculated the position angle of the group members within $R_t$, in order to have a proxy of the direction of the large-scale filament \citep{limousin2013}. Indeed, since galaxy members bear the signature of the hierarchical accretion from surrounding filaments, measuring their spatial distribution provides a neat way to assess the topology of the large scale structures where the main group is embedded and, in turn, the group shape (axial ratios and position angle). We model these shapes of the galaxy member distribution $(R.A.,DEC)$ as an ellipse. The eigenvectors of the covariance matrix $cov(R.A.,DEC)$ represent the principal axes of the ellipse, with the lengths of the principal axes given by the eigenvalues. We found that the galaxy member distribution appears to be elongated, with an axial ratio $1.3\pm0.15$ and position angle $107\pm5$ degrees (measured north through east in celestial coordinates). Errors have been calculated via a bootstrap with resampling approach. The position angle from optical data is roughly in agreement with the X-ray value from the moments of the X-ray distribution (\S\ref{datasets}). %This alignment is expected since the accretion happens preferentially along the cosmic filaments \citep[e.g.,][]{brunino2007}.

The alignment between the optical and X-ray filament is a confirmation of the large-scale structure formation scenario. Indeed, in light of the hierarchical CDM model, the matter is distributed as a network of gigantic filaments (the so-called ``cosmic web''), with their densest regions at the filament nodes and hosting massive clusters of galaxies. Numerical simulations predict that the infall of material into the cluster DM haloes is expected to be preferentially funneled through the filaments where the haloes are embedded, hence leading to an alignment between the major axis of the host haloe, its observables (X-ray, optical) and the large-scale filament \citep{brunino2007,limousin2013}.

\section{X-ray physical properties}\label{phys5346}

In this Section we present the recovered physical properties of NGC~2563. 

We determined the gas inhomogeneities from the $S_X$ map. Thus, we are able to recover gas clumping down to the scales of about 16~arcsec ($\sim5$ kpc). Finally, we apply the method described in the previous Section on $S_X(r)$ to derive the 3D gas density and gas clumpiness. We further performed a spectral (deprojection) analysis (\S\ref{datasets3}). In this step, the task provides directly the 3D temperature radial profile, under the assumptions of sphericity and single-temperature absorbed APEC model.

\begin{figure*}
\begin{center}
% \hbox{
\includegraphics[scale=0.9]{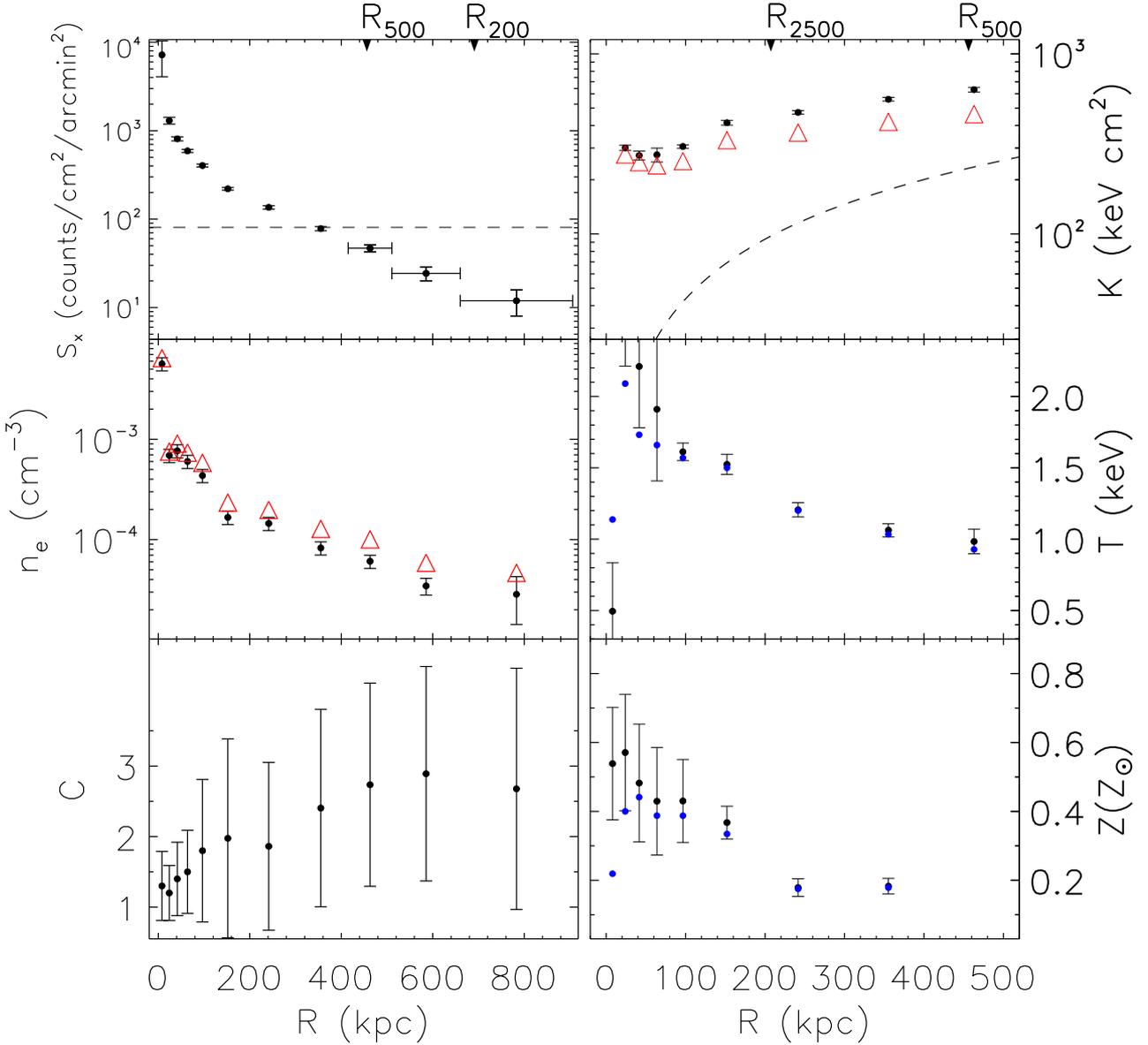}
% \includegraphics[scale=0.39]{ps/kT-S.ps}
% }
\caption{\chandra\ X-ray physical properties of NGC~2563. {\it Upper-left panel:} Exposure-corrected surface brightness profile in the soft X-ray band (0.7-2 keV), obtained by subtraction of the background from the measured signal. The horizontal bars show the width of the annuli (only shown in the last three annuli for clarity). The horizontal bar represents the level of the background. Middle-left panel: 3D gas density profiles. We show results with (circles) and without (triangles) the gas clumping factor correction.  Bottom-left panel: 3D clumping factor profile recovered from the observed two-dimensional intrinsic fluctuations of the X-ray brightness. The average clumping factor for $R\gesssim R_{500}$ is $2.7\pm 0.7$. Top-right panel: radial entropy profile. We show results with (dots) and without (red triangles) the gas clumping factor. The dashed line represents the predictions of \protect\cite{Voit2005b} from pure gravitational collapse, where the entropy is defined as $K(r)=K_{200}\,1.32(r/R_{200})^{1.1}$, $K_{200}$ being a characteristic value of the entropy at an overdensity of 200 \protect\citep[see, e.g., Eq.~2 in][]{Voit2005b}. Middle-right panel: radial temperature profiles. The blue and black dots show the projected and deprojected temperature profiles respectively. For the projected temperature we omitted the errorbars for clarity. Bottom-right panel: metallicity profile (in units of $Z_\odot$, which is the solar abundance of iron). The blue and black dots show the projected and deprojected metallicity profiles, respectively. For the projected metallicity we omitted the errorbars for clarity. From the left to the right, the arrowhead pointers at top of the upper panels indicate $R_{2500}$, $R_{500}$ and $R_{200}$, respectively. $R_{500}=456\pm28$~kpc and $R_{200}=691\pm44$~kpc .}
\label{entps3xkn3fr}
\end{center}
\end{figure*}

The left panel of Figure \ref{entps3xkn3fr} shows the derived 3D clumping factor and gas density profiles for NGC~2563. The gas clumping factor profile becomes larger than the unity outside the core. It then increases steadily out to $R_{500}$, reaching $C \approx 2-3$ in the virialization region. The average clumping factor for $R\gesssim R_{500}$ is $2.7\pm 0.7$. Taking into account the non-negligible statistical uncertainties, these results are in moderate tension with the gas behavior at large radii of other galaxy groups recently reported \citep{humphrey2012,su2015,buote2016} and with the prediction from hydrodynamic numerical simulations \citep{battaglia2013}, which indicate smaller values of gas inhomogeneities. The origin of this discrepancy may be sought in the different ({\em direct} vs. {\em indirect}) methods implemented. In indirect methods (e.g. \citealt{su2015} for a group or \citealt{walker2013} for massive clusters) gas clumping is born out of the comparison between measured and theoretically-predicted entropy profiles \citep{voit2005a}. The observed flattening of the entropy with respect to the power-law behavior from the theory is interpreted only as the result gas inhomogeneities. However, other non-equilibrium physics may also affect the measured entropy, including different temperatures of ions and electrons, and gas (density and temperature) inhomogeneities distribution. We advocate the need to perform a one-to-one comparison between these direct and indirect methods in order to understand the relative impact of these non-equilibrium phenomena. It is unfortunate, however, that we are unable to measure gas temperature, and hence entropy, out to $R_{200}$ in NGC~2563.

As for the discrepancy with simulations, groups contain a wealth of information about the physical processes associated with galaxy formation, including feedback from supernovae, AGN, star formation, or galactic winds. These complex physical properties are only partially understood and captured with modern hydrodynamical simulations. Indeed, systematic differences between hydrodynamic simulations (e.g. smoothed-particle hydrodynamics (SPH) vs. adaptive-mesh refinement (AMR) codes, see \citet{rasia2014}) suggest that the use of these theoretical predictions merits some caution.

As for the gas density slope $\beta=-1/3\,d\log(n_e)/d\log(r)$, our emission measure profile translates into a shallow density profiles, with $\beta \sim 0.42\pm  0.04$ at $R_{500}$ and $\beta \sim 0.44\pm  0.08$ at $R_{200}$, without clumping correction.
%This constant slope is at odds with the steeping observed for massive clusters in previous works by {\em ROSAT} \citep{vikhlinin1999} and from a stacked \chandra\ sample \citep[$\beta \sim 0.940\pm  0.018$ at $R_{200}$][]{morandi2015}. The gas density is also flatter than  the value measured by \cite{wong2016} on the Antlia Cluster ($\beta=0.58^{+0.05}_{-0.05}$, and flatter than the density slope of 1.65-2.25 at $R_{500}$ measured with 43 nearby galaxy groups using \chandra\ \citep{sun2009}. However, it agrees with {\em XMM} measurements of the Virgo cluster \citep[$\beta=0.40 \pm 0.04$,][]{urban2011}, which has a low temperature as well ($kT\sim 2.3$ keV). When corrected for clumping, the density profile is only slightly steeper ($\beta \sim 0.46$ for  $R\gesssim R_{500}$).
This constant slope is significantly smaller than those observed for massive clusters in previous works by {\em ROSAT} \citep{vikhlinin1999} and from a stacked \chandra\ sample \citep[$\beta \sim 0.940\pm  0.018$ at $R_{200}$,][]{morandi2015}. For the Antlia Cluster ($kT_{500} \sim 1.5$ keV), \cite{wong2016} fit the density profile from 0.2 $R_{200}$ to 1.3 $R_{200}$ with a power law model and the corresponding $\beta$ is about 0.58. For both groups, the density slope at  $R_{500}$ is consistent with the previous results from 22 groups with the \chandra\ data \citep{sun2009}, where there was a hint of smaller $\beta$ for $kT$ $<$ 1.4 keV.  When corrected for clumping, the density profile of NGC~2563 is only slightly steeper ($\beta \sim 0.46$ for  $R\gesssim R_{500}$). This shallow slope is consistent with a scenario where much of the low-entropy gas in poor groups has been ejected to large radii by strong feedback \citep[e.g.][]{mccarthy2011}.

The right panel of Figure \ref{entps3xkn3fr} shows the 3D temperature profile of NGC~2563, which shows a drop by roughly a factor of 2 from the peak temperature at $R\approx 0.7\,R_{200}$. The profile appears to be regular, with a low-temperature in the center, hinting the presence of a cool-core. The temperature profile goes down to $kT\lesssim 1$ keV at the X-ray boundary. 

Our derived entropy profile tends to flatten at large radii ($r\gesssim R_{500}$), being larger than the prediction of adiabatic collapse $K_{adi}$ \citep{voit2005a}. For example, $K_{adi,500}\simeq156$ keV cm$^{-2}$, a few times lower than our measurements. Entropy values at $R_{2500}$ and $R_{500}$ are 366$\pm$10 keV cm$^{2}$ and 532$\pm$28 keV cm$^{2}$ respectively, which falls on the $K - T$ relations reported by \citet{sun2009}. Thus, the entropy excess observed for NGC~2563 is typical for groups with similar mass. After the gas clumping correction, the entropy profile is still flat, with a slope of $0.45 \pm 0.20$, which clearly deviates from 1.1 of the baseline adiabatic model. Such a flat entropy profiles may indicate a large impact of non-gravitational effect (e.g., cooling, preheating and AGN feedback) are very pronounced. This impact is expected to be larger in groups than massive galaxy clusters at least in the internal regions ($\lesssim R_{500}$). While we are unable to measure the entropy in the virialization region $R_{500}-R_{200}$, recent results from \cite{su2015} on the fossil cluster RXJ1159+5531 suggest that $K$ is consistent at $R_{200}$ with the predictions from gravity-only simulations.

We point out that our entropy profiles may be biased upwards by the inability of the X-ray observatory to probe gas at very low temperatures. By assuming a log-normal distribution of the gas (density and temperature) inhomogeneities, for low temperatures $\sim T_{\rm cold}/(1+z)\lesssim 0.25$ keV the measured gas density would be biased downwards and hence the entropy upwards. On the one hand, this is due to the fact that most of the emission captured by the X-ray telescope arises from the warmer phase of the hIGM, with temperature $\gesssim0.6(1+z)$ keV. On the contrary, most the emission of the cold phase would be at very low energies, where the \chandra\ effective area is small, and related to emission lines rather than bremsstrahlung. These effects would stronger for groups than massive clusters, given the larger emission at low temperatures and larger contribution of emission lines over bremsstrahlung. The incorrect assumption that the gas is at a single-temperature would exacerbate the bias on the gas density. On the other hand, the implemented single-temperature APEC model may bias downwards the X-ray temperature, given the contaminating low-temperature tail of the temperature distribution.

Finally, we discuss the metal abundance in the hIGM. \chandra\ data sets measure the metallicity structure of the intra-cluster gas with high precision and spatial resolution roughly out to $0.6\,R_{200}$. Our results show that the group outskirts are also substantially metal-enriched, to a level amounting to approximately 0.2 of the Solar metallicity (see Figure \ref{entps3xkn3fr}). Our data provide direct observational evidence that the hIGM is enriched by metals out to $0.5\,R_{200}$. The average iron abundance is also consistent with previous measurements for 39 groups from \cite{sun2012}, after the difference on AtomDB and abundance table is accounted for. The hIGM metal content is a key observable to infer the cumulative past star formation history in galaxy clusters, allowing to study the enrichment processes. While the injection of metals is tied to processes of star formation, its radial profile is set by different physical processes, such ram-pressure stripping, galactic winds triggered by supernovae and AGN activity, and merger mechanism \citep[e.g.,][]{gnedin1998}. As a proxy of feedback processes due to star formation, which release energy into the hIGM, this non-negligible abundance in the outer regions is in agreement with the entropy excess with respect to models of pure gravitational collapse that we previously discussed.

\section{Gas and baryon fraction}
\begin{figure}
\begin{center}
\includegraphics[scale=0.45]{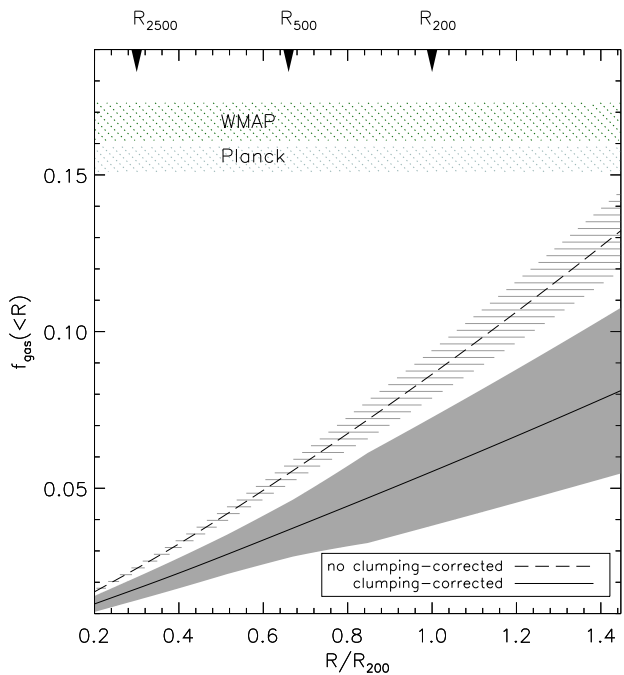}
\caption{Cumulative gas fraction $f_{\rm gas}(<R)$ as a function of the radius (normalized to $R_{200}$). The solid (dashed) lines represent the values by correcting for the gas clumping (without clumping correction), while the 1-$\sigma$ errors are represented by the gray shaded region (hatched region). The two horizontal dotted regions represent, from the top (in dark green) to the bottom (in grey), the 1-$\sigma$ confidence level on the primordial baryon fraction from {\em WMAP}-9 years ($\Omega_b/\Omega_m=0.167\pm0.006$) and {\em Planck} ($\Omega_b/\Omega_m=0.155\pm0.004$), respectively. We interpolated the data via a least-squares constrained spline approximation so as to have continuous functions and thus improve the readability of the figures. $f_{\rm gas}(<R)$ scales with the Hubble constant as $f_{\rm gas}\propto h^{-3/2}$.}
\label{entps3xkn3frsbvf}
\end{center}
\end{figure}

The enclosed hot gas fraction $f_{\rm gas}(<R) =M_{\rm gas}(<R)/M_{\rm tot}(<R)$ is presented in Figure \ref{entps3xkn3frsbvf}. We present X-ray measurements of the hot gas fraction out to $R_{100}$. We also discuss the impact of dynamical state, (multi)-temperature cluster, and asphericity on our results. In particular, we investigated the biases on our gas fraction measurements due to the effects of inhomogeneities of the gas distribution and non-thermal pressure. Clearly, $f_{\rm gas}$ is inconsistent with the cosmic value. The hot gas fraction at $R_{100}$ without and with clumping correction is $\sim$ 0.13 and 0.07 respectively, with clumpiness biasing upwards the gas mass fraction of about $C^{0.5}\sim 1.6$ \citep{roncarelli2013,morandi2013b,morandi2014}.

Our measurements are also affected by violation of HE. We remember that $R_{200}$ and $R_{100}$ have been statistically determined by assuming an NFW distribution for the total matter, with a value of the concentration parameter from N-body simulations. We do not infer $R_{500}$ by applying HE by means of gas density and deprojected temperature, since we are unable to measure the latter quantity out to $R_{200}$; neither we measure the baryon budget by assuming HE in the regions beyond $R_{500}$. Rather, we anchor our results to the determination of $R_{500}$ via the HE assumption and by exploiting the existence of low-scatter scaling relations between the total thermal energy $Y_X$ and the total mass. While our results hinge on these assumption on total mass distribution, this method adds the benefits to straightforwardly investigate e.g. biases due to the assumption of HE at $R_{500}$, which are discussed in the literature. For example, if we define the mass bias $b$ between the true and hydrostatic mass $M_{hydro}$ ($M_{true}=M_{hydro}/(1-b)$), even a modest bias $b\sim 0.1$ translates into a bias (upwards) of $\sim 14\%$ of the gas fraction at $R_{200}$.
The amount of mass bias has been recently debated, with some works that constrain it in the range 0.2 --- 0.45 at $\sim$ $R_{500}$ \citep[e.g.,][]{okabe2013,vonderlinden2014,hoekstra2015}; other recent works, however, disfavor strong departure from HE \citep[e.g.][]{israel2014,applegate2014}. Moreover, mass bias is still poorly understood in the group scale. Initial results recently \citep{kettula2013,kettula2015,lieu2016} suggest a comparable mass bias in groups compared to clusters. However, the number of groups with weak lensing mass is quite small and the existing scaling relation may be subject to various selection biases.
Results from simulations also suggest that the level of non-thermal pressure is insufficient \citep[$\sim$11\%,][]{nagai2007a,nelson2012,nelson2014} to reconcile the aforementioned inconsistencies. With this large HE bias, our measured gas fraction would be even much more biased upwards.

Gas fractions in simulated groups and clusters are directly related to the strength of cooling and star formation \cite[e.g.][]{kravtsov2005}, and to the way in which energetic feedback from the central SMBH alters the density distribution of the ICM \citep[e.g.][]{puchwein2008,bower2008,fabjan2010,mccarthy2010,mccarthy2011,kravtsov2012}. Interestingly, hot gas content from galaxies to clusters can also be studied by stacking the {\em Planck} data \citep{planck2013,greco2015}. As shown by \cite{lebrun2015}, the {\em Planck} results may suggest a large amount of hot gas has been ejected out of $R_{500}$ in low mass systems. The results presented in this work are consistent with this picture, given the large deficict of baryons with respect to the primordial baryonic budget.

There is not any measurement of the stellar mass in NGC~2563. As a nearby system, even such a measurement can in principle be done with the {\em SDSS}, {\em 2MASS} and {\em WISE} data, a potential key stellar mass component, intragroup light, is missing \citep{mcgee2010,gonzalez2013}. Thus, we rely on established scaling relations to estimate the stellar mass and baryon fraction in NGC~2563. However, the recent results from different works \citep{leauthaud2012,lin2012,giodini2012,budzynski2014,chiu2016} vary by a factor of nearly 3 for the mass of NGC~2563, $f_{\textrm{stars}}\sim 0.012 -0.035$. If we extrapolate the \cite{gonzalez2013} relation to the mass of NGC~2563, the stellar fraction is even greater, $\sim$ 0.06. Nevertheless, the baryon fraction within $R_{500}$ is unlikely to reach the universal value, given the potentially important clumping correction and the likely significant mass bias, unless the intragroup light can contribute to a large fraction of the group baryons.

Finally, we analyzed the impact of the $M_{500}-Y_{X,500}$ relation from \cite{sun2009} used to determine the boundary radius $R_{500}$. Note that our mass profile is in good agreement with the mass recovered by assuming HE for $R_{2500}\lesssim R\lesssim R_{500}$ (Fig .\ref{evderyddd2sdvbr6}). We point out, however, that if we used instead the $M_{500}-kT$ relation from \cite{sun2009}, we would have masses roughly 50\% larger than those presented here. This may occurr if the slope of the $M_{500}-kT$ relation  steepens in the mass range of groups, and it would translate into a comparable bias (downwards) on the gas fraction for $R\gesssim R_{500}$.

\section{Conclusions}\label{conclusion33}
Groups are systems where baryonic physics (e.g., cooling, preheating and AGN feedback) begins to dominate over gravity \citep[e.g.][]{voit2005a,sun2012}. Thus, they represent a unique opportunity to study the baryonic physics, since its impact is more significant in low-mass systems with respect to massive clusters. 

In this paper, we present an analysis of the hIGM physical properties, namely gas density and gas clumping, gas temperature distribution, entropy and gas fraction, of the galaxy group NGC~2563. This is a group with global temperature of about 1.2 keV, a redshift $z=0.0157$. This system is characterized by mosaic \chandra\ observations to reach beyond $R_{200}$, allowing the virialization region to be studied.

We measure the spectroscopic physical properties (gas temperature, entropy and metal content) out to about $R_{500}$. The measure entropy profile - with a flat slope and higher normalization with respect to the adiabatic model predictions - indicate a large impact of non-gravitational effects atop of the gravitational energy. This is confirmed by our metallicity measurements, where the abundance amounts to approximately 0.2 of the Solar metallicity out to $0.5\,R_{200}$. As a key observable to infer the cumulative past star formation history in galaxy clusters, this significant value of the metal content retains the fingerprint of the feedback processes due to star formation, cooling, galactic winds and AGNs, which release energy into the hIGM, and shape its physical properties accordingly.
              
We then measured the baryon content of the galaxy group (gas density and slope, gas clumping and gas fraction). We applied a direct method to recover the gas density inhomogeneities. We found a large value of the gas clumping factor, which reaches $C \approx 2-3$ in the virialization region, although with non-negligible statistical uncertainties. The average clumping factor for $R\gesssim R_{500}$ is $2.7\pm 0.7$. These results are different from the gas behavior at large scales recently reported in the study of other galaxy groups \citep{humphrey2012,su2015,buote2016} and with the prediction from hydrodynamic numerical simulations \citep{battaglia2013}, which indicate smaller values of gas inhomogeneities. 

The cumulative gas fraction (clumping-corrected) reaches 0.07, well below the value from the primordial baryon fraction. This large discrepancy may suggest that a large amount of hot gas has been ejected beyond the virialization region in low mass systems, as indicated by recent {\em Planck} results \citep{lebrun2015}. This confirm the large impact of non-gravitation processes in the outer volumes, which leave its fingerprint on the density distribution of the ICM. By studying the latter, we can thus unearth the 'fossil record' of these non-gravitational processes and the ongoing large-scale structures formation scenario. 

We stress that this is the first observational measurement of the physical properties of poor groups beyond $R_{500}$. Our results therefore can provide an anchor for numerical models of ICM physics and for simulations of the formation and ongoing growth of galaxy clusters and groups. In particular, understanding the baryon content and physics has both astrophysical and cosmological ramifications. As for the cosmology, the effects of baryonic physics are not large but still non-negligible in massive clusters, also near $R_{500}$, which is widely used as a benchmark boundary e.g. to infer the cluster mass for cosmological purposes \citep{vikhlinin2009c,morandi2016}. Therefore, these non-gravitational processes need to be calibrated to use clusters if we aim at using clusters as precise cosmology probes. As for the astrophysics, the same baryonic physics determines galaxy formation and evolution, and the amount of energy injection from feedback, heating and cooling of gas, which have to be tuned to preserve the cosmic stellar fraction, the galaxy luminosity function, and the observed ICM properties.

\section*{acknowledgements}
A. M. and M. S. acknowledge support from \chandra\ grant GO4-15115X and NASA grants NNX14AI29G and NNX15AJ30G. We thank the anonymous referee for the careful reading of the manuscript and suggestions, which improved the presentation of our work.

\begin{appendix}

\section{Spectral temperatures dependency on atomic database and absorption}\label{app1}

\begin{figure}
\begin{center}
\includegraphics[scale=0.45]{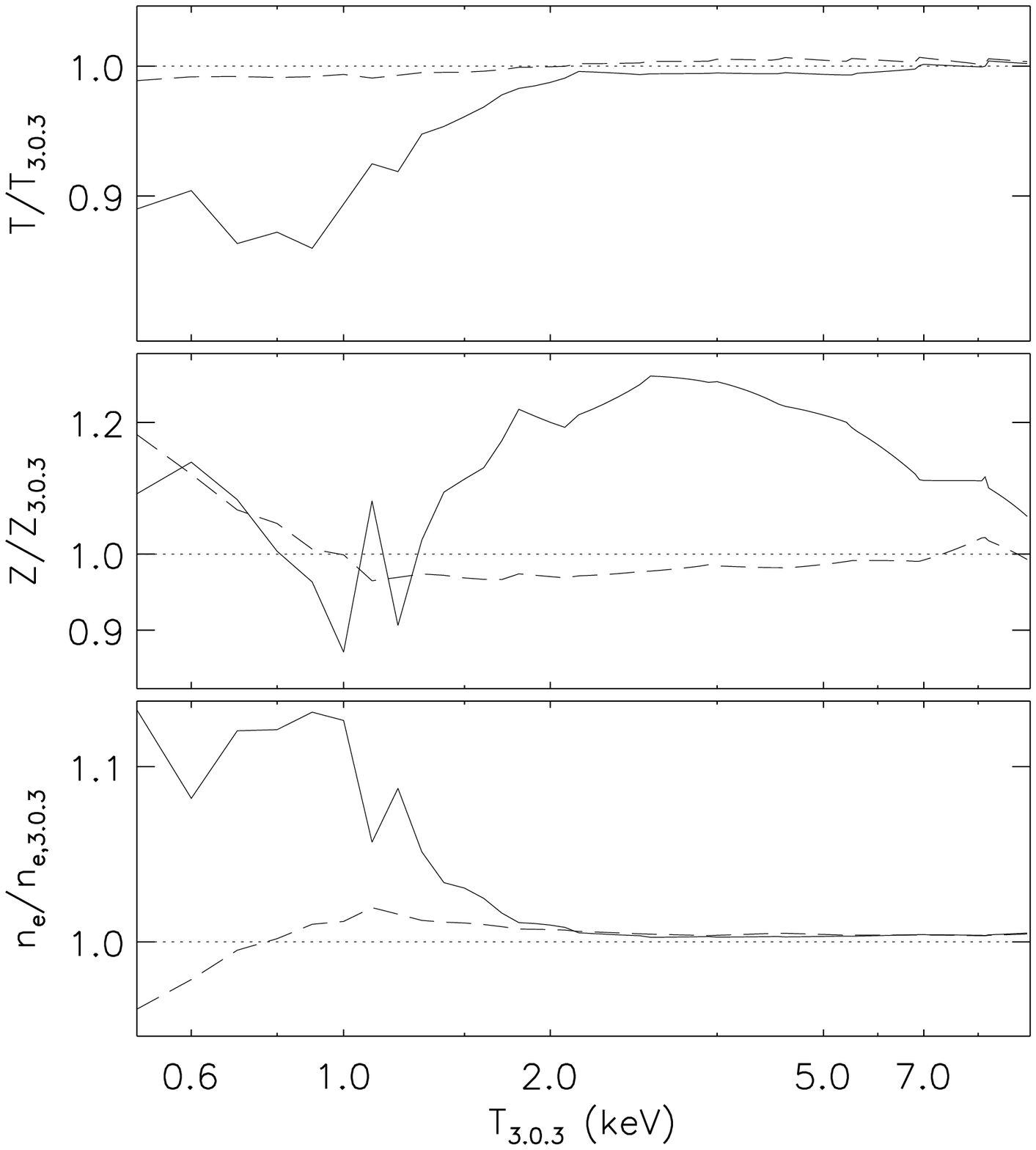}
\caption{Comparison between the spectral temperatures, metallicities and electron densities recovered via the AtomDB databases 2.0.2 (dashed line) and 1.3.1 (solid line) with respect to the AtomDB databases 3.0.3.}
\label{evderyddd2}
\end{center}
\end{figure}

We present a comparison between spectral temperatures and metallicities recovered via the AtomDB databases 2.0.2 and 1.3.1 with respect to the AtomDB databases 3.0.3. We performed mock simulations of a single-temperature plasma at temperatures $T_{ 3.0.3}$, $Z=0.3\, Z_{\odot}$ and hydrogen column density $N_H=4.05\times 10^{20}{\mbox{cm}}^{-2}$ for the aforementioned atomic databases. Note that in our work we adopted the APEC emissivity model \citep{foster2012} and the AtomDB (version 3.0.3) database of atomic data, and we employed the solar abundance ratios from \cite{asplund2009}.

For each temperature, the mock simulations of an absorbed spectra with the AtomDB databases 3.0.3 have been then fitted by implementing different atomic datasets (2.0.2 and 1.3.1). Free parameters in the spectral fit are temperature, metal abundance, normalization. We fixed the hydrogen column density $N_H$ to the Galactic value by using the Leiden/Argentine/Bonn (LAB) HI-survey \citep{kalberla2005}. The redshift was frozen to the value obtained from optical spectroscopy. The difference between atomic databases 2.0.2 and 3.0.3 is, in general, within a few percent, increasing at low temperature ($kT\lesssim 1$ keV), in particular with respect to the metallicity. The difference between the atomic databases 1.3.1 and 3.0.3 is appreciably larger, leading to a bias downwards (upwards) on the recovered spectral temperature (electron density) of $\sim 10\%$ at $kT\lesssim 3$ keV. These results are in agreement with previous findings \citep{lovisari2015}.

\section{Systematic uncertainties}
Here we discuss the bias related to our choice to fix the hydrogen column density $N_{\rm H}$ to the Galactic value ($N_{\rm H}=4.05\times 10^{20}{\mbox{cm}}^{-2}$), which has been determined via the Leiden/Argentine/Bonn (LAB) HI-survey \citep{kalberla2005}. Indeed, the aforementioned radio survey measures the hydrogen column density from the HI-21 cm line, only providing the neutral hydrogen along the line of sight. In this way, however, the molecular and ionized gas components are not accounted for. Although the neutral hydrogen usually contributes most of the total hydrogen for $N_{\rm H}\lesssim 10^{21} {\mbox{cm}}^{-2}$, our assumed value of $N_{\rm H}$ might potentially underestimate the X-ray absorption. \cite{willingale2013} provided a method to account for the molecular hydrogen by means of the dust extinction in the B and V bands, and via X-ray afterglows of gamma ray bursts. Thus, by fixing the absorption to the the value of the hydrogen column density as determined by \cite{willingale2013} ($N_{\rm H}=4.67\times 10^{20}{\mbox{cm}}^{-2}$) we estimated a bias downwards (upwards) of the temperature (metallicity) of $\lesssim$ 2\%. If we leave the absorption as a free parameter in the spectral fit, we obtain consistent values of temperature and metallicity with respect to our baseline spectral fit, 4.3$^{+2.1}_{-1.3} \times 10^{20}{\mbox{cm}}^{-2}$.

\section{Hydrostatic mass determination}\label{appe461r23rfc}
\begin{figure}
\begin{center}
\includegraphics[scale=0.45]{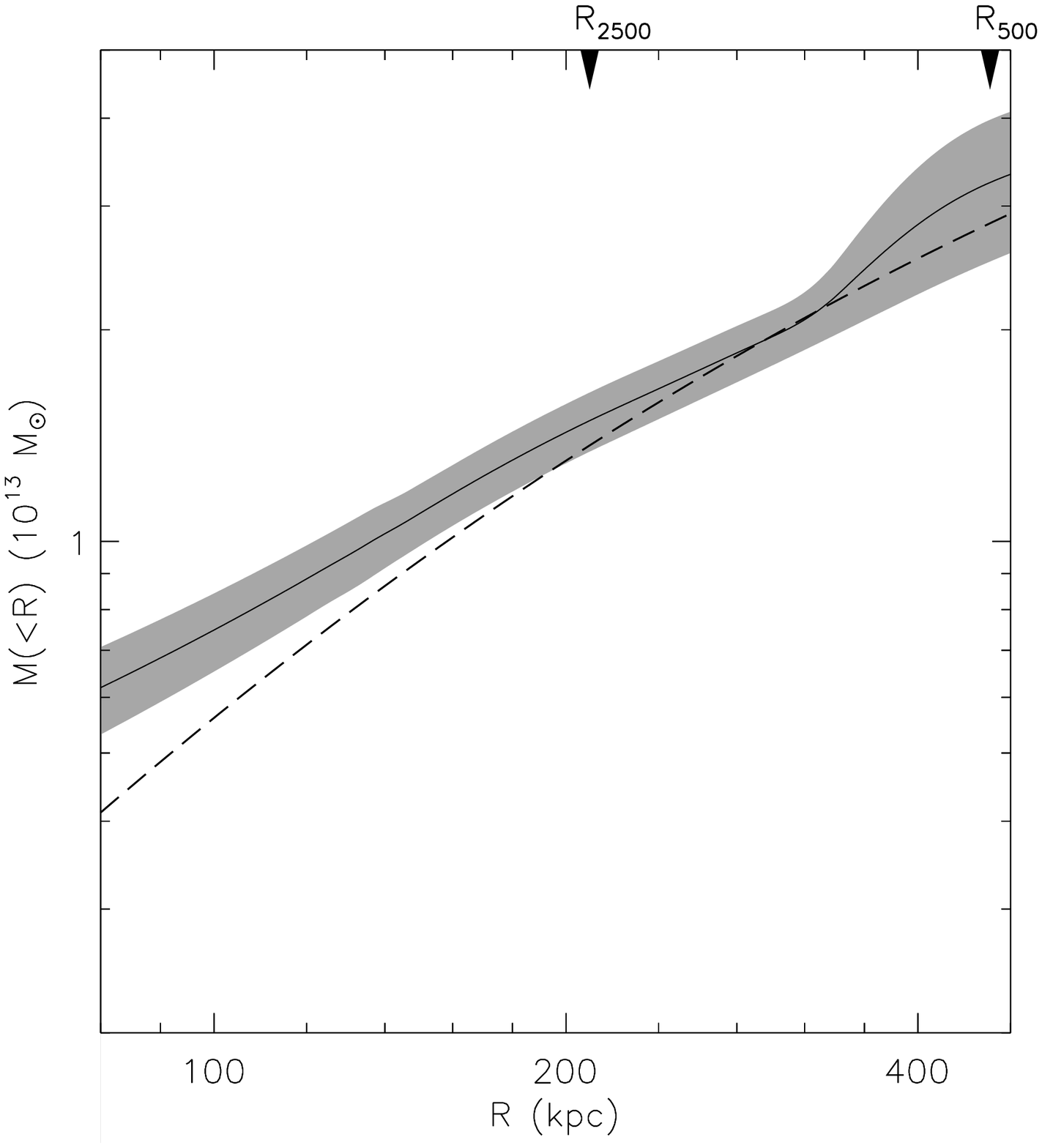}
\caption{Comparison between the cumulative HE mass (solid line with the 1-$\sigma$ errors represented by the gray shaded region) and the mass (dashed line) recovered by assuming: i) the $M_{500}-Y_{X,500}$ relation of \protect\cite{sun2009}; ii) an NFW distribution with parameters $(c,R_{200})$, with a value of the concentration parameter via the $c-M_{\rm vir}$ relation from N-body simulations (\S\ref{datasets3}). In order to improve the readability of the figure, we fitted our X-ray measurements via the analytical functions of gas density and temperature presented in \protect\cite{vikhlinin2006}, so as to have a continuous HE mass profile. The errorbars on the mass profile from the $M_{500}-Y_{X,500}$ relation have been omitted in the figure for clarity.}
\label{evderyddd2sdvbr6}
\end{center}
\end{figure}

In this Section we present the mass recovered by assuming HE. We used the gas (without clumping correction) density and gas temperature profiles and apply the HE equation to calculate the enclosed cumulative total mass. The mass profile is compared to the our baseline mass determination outlined in \S\ref{datasets3}, where we assumed: i) the $M_{500}-Y_{X,500}$ relation of \protect\cite{sun2009}; ii) an NFW distribution with parameters $(c,R_{200})$, with a value of the concentration parameter via the $c-M_{\rm vir}$ relation of \cite{dutton2014} based on the results from N-body simulations (\S\ref{datasets3}). Note that there is a remarkable good agreement between the two mass estimates for $R_{2500}\lesssim R\lesssim R_{500}$ (Fig .\ref{evderyddd2sdvbr6}).

\end{appendix}

% \bibliographystyle{mnras}
% \bibliography{master}

\newcommand{\noopsort}[1]{}

\end{document}